\documentclass[preprint,3p,a4paper,sort&compress]{elsarticle}
\makeatletter
\def\ps@pprintTitle{%
	\let\@oddhead\@empty
	\let\@evenhead\@empty
	\def\@oddfoot{\centerline{\thepage}}%
	\let\@evenfoot\@oddfoot}
\makeatother
\usepackage[final]{changes}
\definechangesauthor[name={Ref1}, color=red]{r1}
\definechangesauthor[name={Ref2}, color=orange]{r2}

\usepackage{graphicx}
\usepackage{epstopdf}
\usepackage{dcolumn}
\usepackage{amsmath}
\usepackage{mathptmx, courier, pifont}
\usepackage[scaled=0.92]{helvet}
\usepackage[T1]{fontenc}
\usepackage{textcomp}
\usepackage{color}
\usepackage{lineno,hyperref}
\modulolinenumbers[1]

\journal{Nuclear Instruments and Methods in Physics Research Section A}

\usepackage{subfigure,dcolumn}
\begin{document}
	
	\title{Characterization of a double Time-Of-Flight detector system for accurate velocity measurement in a storage ring using laser beams}
	
	\cortext[mycorrespondingauthor]{Corresponding authors}
	\author[A,B]{Xin-Liang Yan\corref{mycorrespondingauthor}}
	\ead{yanxinliang08@impcas.ac.cn}
	\author[A,C]{Rui-Jiu Chen}
	\author[A]{Meng Wang\corref{mycorrespondingauthor}}
	\ead{wangm@impcas.ac.cn}
	\author[A]{You-Jin Yuan}
	\author[A]{Jian-Dong Yuan}
	\author[A]{Shao-Ming Wang}
	\author[A]{Guo-Zhu Cai}
	\author[A,D]{Min~Zhang}
	\author[A]{Zi-Wei Lu}
	\author[A,D]{Chao-Yi Fu}
	\author[A,D]{Xu Zhou}
	\author[A]{Dong-Mei Zhao}
	\author[C]{Yuri A. Litvinov}
	\author[A]{Yu-Hu Zhang}
	\address[A]{Institute of Modern Physics, Chinese Academy of Sciences, 509 Nanchang Rd., Lanzhou 730000, China}
	\address[B]{Argonne National Laboratory,
		9700 S. Cass Avenue Lemont, Illinois 60439, USA}
	\address[C]{GSI Helmholtzzentrum f\"{u}r Schwerionenforschung, Planckstra\ss{}e 1, 64291 Darmstadt, Germany}
	\address[D]{University of Chinese Academy of Sciences, No.19(A) Yuquan Road, Shijingshan District, Beijing  100049, China}
	\begin{abstract}
		The Isochronous Mass Spectrometry (IMS) is a powerful tool for mass measurements of exotic nuclei with half-lives as short as several tens of micro-seconds in storage rings. In order to improve the mass resolving power while preserving the acceptance of the storage ring, the IMS with two Time-Of-Flight (TOF)
		detectors has been implemented at the storage ring CSRe in Lanzhou, China. Additional velocity information beside the revolution time in the ring can be obtained for each of the stored ions by using the double TOF detector system. In this paper, we introduced a new method of using a 658 nm laser range finder and a short-pulsed ultra-violet laser to directly measure the distance and time delay difference between the two TOF detectors which were installed inside the $10^{-11}$ mbar vacuum chambers. The results showed that the distance between the two ultra-thin carbon foils of the two TOF detectors was ranging from 18032.5 mm to 18035.0 mm over a measurable area of 20$\times$20 mm$^2$.
		Given the measured distance, the time delay difference which comes with signal cable length difference between the two TOF detectors was measured to be $\Delta t_{delay1-2}=99$(26) ps. The new method has enabled us to use the speed of light in vacuum to calibrate the velocity of stored ions in the ring. The velocity resolution of the current double TOF detector system at CSRe was deduced to be $\sigma(v)/v=4.4\times 10^{-4}$ for laser light, mainly limited by the time resolution of the TOF detectors. 
	\end{abstract}
	
	
	\maketitle
	
	\section{Background}\label{sec.1}
	\begin{figure}[htb!]
		\begin{center}
			\includegraphics[width=.9\hsize]{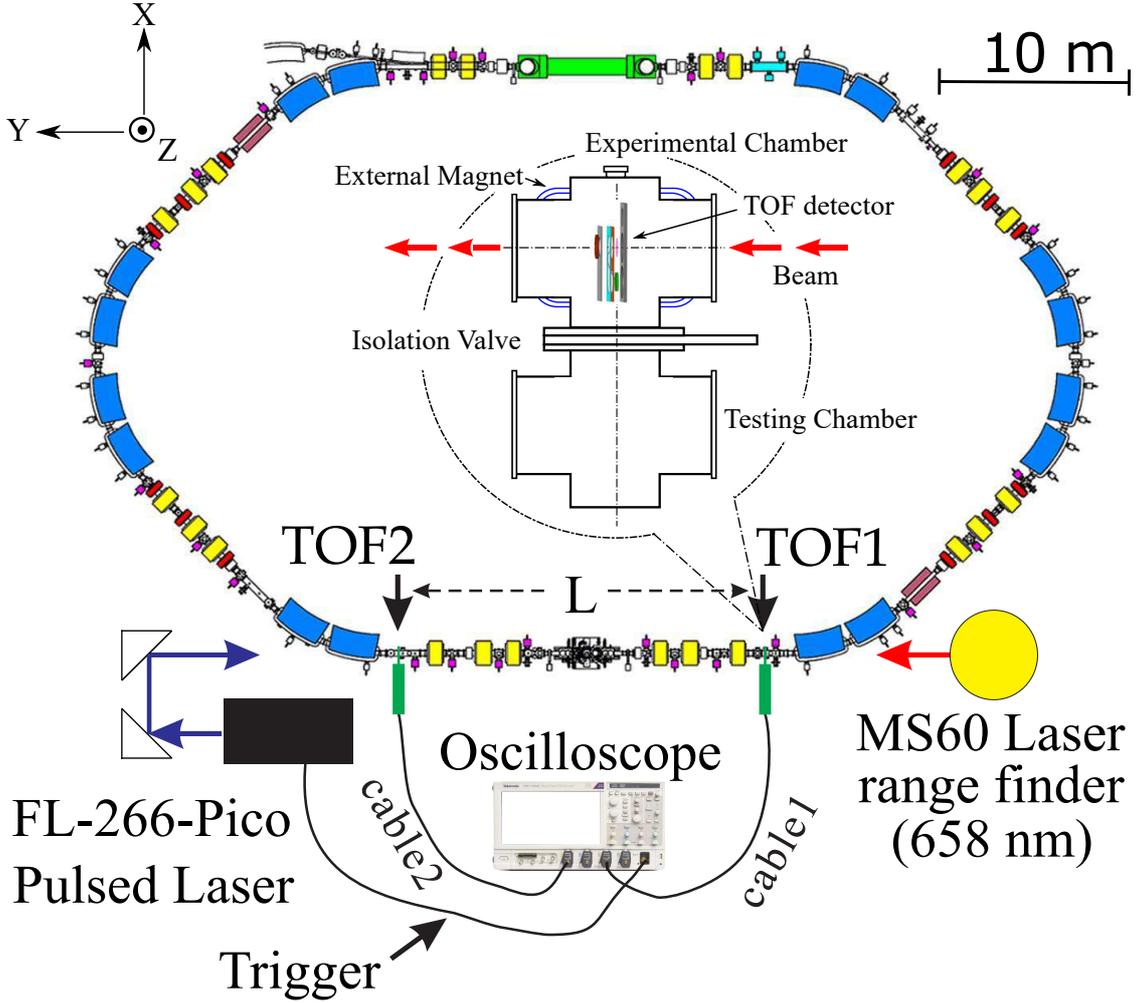}
		\end{center}
		\caption{(Color online) Schematic view of the storage ring CSRe. The dipole and quadruple magnets are indicated as blue and yellow rectangles, respectively. Two Time-Of-Flight detectors (indicated as TOF1 and TOF2) are installed inside the vacuum chamber at one of the straight sections of CSRe. In order to let the laser beams to shine through the beam line, two quartz vacuum windows \cite{mpf} were installed on both ends of the vacuum chamber of the beam pipe. The TOF detectors are movable in X-direction between \replaced{testing }{tesing }chamber and experimental chamber.}
		\label{fig1}
	\end{figure}
	
	The accurate knowledge of the masses of atomic nuclei is particularly important for nuclear structure and nuclear astrophysics \cite{Lunney2003,Blaum2006}.
	The challenge today is to obtain accurate masses of nuclei located far away from the valley of $\beta$-stability \cite{100yms}.
	However, such nuclei are short-lived and are produced with small cross section.
	Therefore, highly efficient and fast measurement techniques are required.
	The Isochronous Mass Spectrometry (IMS) is one of such techniques
	which is realized at in-flight separation radioactive ion beam facilities \cite{Litvinov2013,Bosch2013,ZhangPS2016}.
	The IMS experiments are performed today at three heavy-ion storage ring facilities,
	namely the experimental storage ring ESR at GSI Helmholtz Center in Darmstadt, Germany
	\cite{wollnik1987,tro1992,hausmann2000,Hausmann2001, Stadlmann2004, Sun2008,franzke08, KnobelEPJ2016,SunPLB2010,KnobelPLB2016}, 
	the experimental cooler-storage ring CSRe at the Institute of Modern Physics in Lanzhou, China
	\cite{Meng2009,TuNIMA2011,TuPL2011,ZhangPRL2012,YanAPL2013,ShuaiPLB2014,Xu2016,Zhang2017,Xing2018,fu18,zhang18},
	and the rare-ion storage ring R3 at the RIKEN Nishina Center in Tokyo, Japan
	\cite{Yamaguchi08,Ozawa2012,Yamaguchi2013a,Yamaguchi2013b,Wakasugi2015,Yamaguchi2015a,Yamaguchi2015b}.
	
	With the IMS masses of exotic nuclei can be deduced from precise revolution time measurements of the ions circulating in the isochronous storage ring \cite{wollnik1987}. The revolution times, which are measured by a Time-Of-Flight (TOF) detector \cite{tro1992,hausmann2000,Hausmann2001,tu2010,diw2015,kuz2016,mei,zhang14}, are related to the mass-to-charge ratios ($m/q$) of the stored ions in the ring \cite{franzke08}:
	\begin{equation}
	\frac{\Delta T}{T} = \frac{1}{\gamma^2}\frac{\Delta (m/q)}{m/q} -\left(\frac{1}{\gamma^2}-\frac{1}{\gamma_t^2}\right)\frac{\Delta (B\rho)}{B\rho},
	\label{IMSformula}
	\end{equation} 
	where $\gamma$ is the relativistic Lorentz factor, $B\rho=\frac{m}{q}\gamma v$ is the magnetic rigidity and $v$ is velocity of the ions. The transition energy factor $\gamma_t$ is a machine-parameter determined by the ion optics of the ring and is defined as $\gamma_t^2=[\Delta (B\rho)/(B\rho))]/(\Delta C/C)$ where $C=vT$ is the orbital length of the stored ions. In the isochronous condition of the ring, the $\gamma_t$ is set equal to the $\gamma$ of the target ion so that the second term of Eq.(\ref{IMSformula}) is negligible. 
	In reality, $\gamma_t$ of a storage ring is not a constant, but is a function of the $B\rho$ or $C$ of the stored ions. Moreover, even for the target nuclei,  all individual ions are injected at different $\gamma$ which do not match perfectly $\gamma_t$, therefore the spread of the revolution times $\sigma(T)$ is not zero and the mass resolving power is a finite value. The achievable mass resolving power $R$($1\sigma$-value) of IMS is limited by the $B\rho$-acceptance of the storage ring:
	\begin{equation}
	R=\frac{m}{\Delta m}\approx\frac{1}{\gamma^2}\frac{T}{\sigma(T)}=\frac{\gamma_t^2}{\gamma^2-\gamma_t^2}\frac{B\rho}{\sigma(B\rho)}=\frac{\gamma_t^2\gamma^{-2}}{\gamma^2-\gamma_t^2}\frac{v}{\sigma(v)}.
	\label{IMSBrho}
	\end{equation} 
	Typical $B\rho$-acceptance $\sigma(B\rho)/(B\rho)$ of the CSRe storage ring was about $8.5\times10^{-4}$ for the isochronous setting at $\gamma_t=1.39$ \cite{jxia}. Obviously, $R$ has the largest value for the target ion. For those ion species other than the target ion, $R$ quickly drops down because $\gamma$-value deviates significantly from $\gamma_t$-value due to different $m/q$ of the ions. 
	
	A novel way to increase $R$ without reducing the acceptance of the ring was first proposed at GSI in 2005 \cite{ILIMA2006,Geissel2005}. The idea is to measure the velocity for each stored ions with two TOF detectors in the ring such that the revolution time spread can be reduced by using the velocity information in the off-line data analysis \cite{Geissel06,dolinski,xu15,shuai16}. 
	In this way, the resulting mass resolving power can be improved for all the stored ions over a large $m/q$-range and the systematic error caused by the second term in Eq.(\ref{IMSformula}) can be reduced in the mass calibration of the revolution time spectrum.
	The IMS with two TOF detectors has been set up at CSRe and on-line test experiments have been carried out recently \cite{xing2015}. Two new TOF detectors \cite{tu2010,mei,zhang14} have been installed in the straight section of the ring (see Fig. \ref{fig1}), new ion optics of the ring with zero dispersion at the two TOF detectors and proper $\gamma_t$-value and $B\rho$-acceptance have been established for CSRe \cite{ge18}.  Note that with the velocity information obtained in IMS, the $\gamma_t$ of the ring as a function of orbital length has been deduced in a straightforward way \cite{chen18,ge18}. Therefore, with both $v$ and $T$ measured, the achievable mass resolving power $R$ will no longer be limited by the physical acceptance of the ring but by the precision of the velocity measurement of each stored ion (see last term in Eq. \ref{IMSBrho}).
	
	In this paper, we have developed a new method employing laser beams to study the characteristics of the double TOF detector system in CSRe. The distance between the two TOF detectors was measured using a laser range finder and the time delay difference which comes with signal cable length difference between the two TOF detectors was measured with help of a picosecond-pulsed ultra-violet (UV) laser. The velocity-resolution of the double TOF detector system and time-resolution of one TOF detector were determined. The lasers, which can be used \emph{in situ and operando} during the IMS experiment at CSRe, have enabled us to use the speed of light in vacuum to calibrate the velocity of heavy ions.
	\begin{figure}[htb]
		\begin{center}
			\includegraphics[width=.42\hsize]{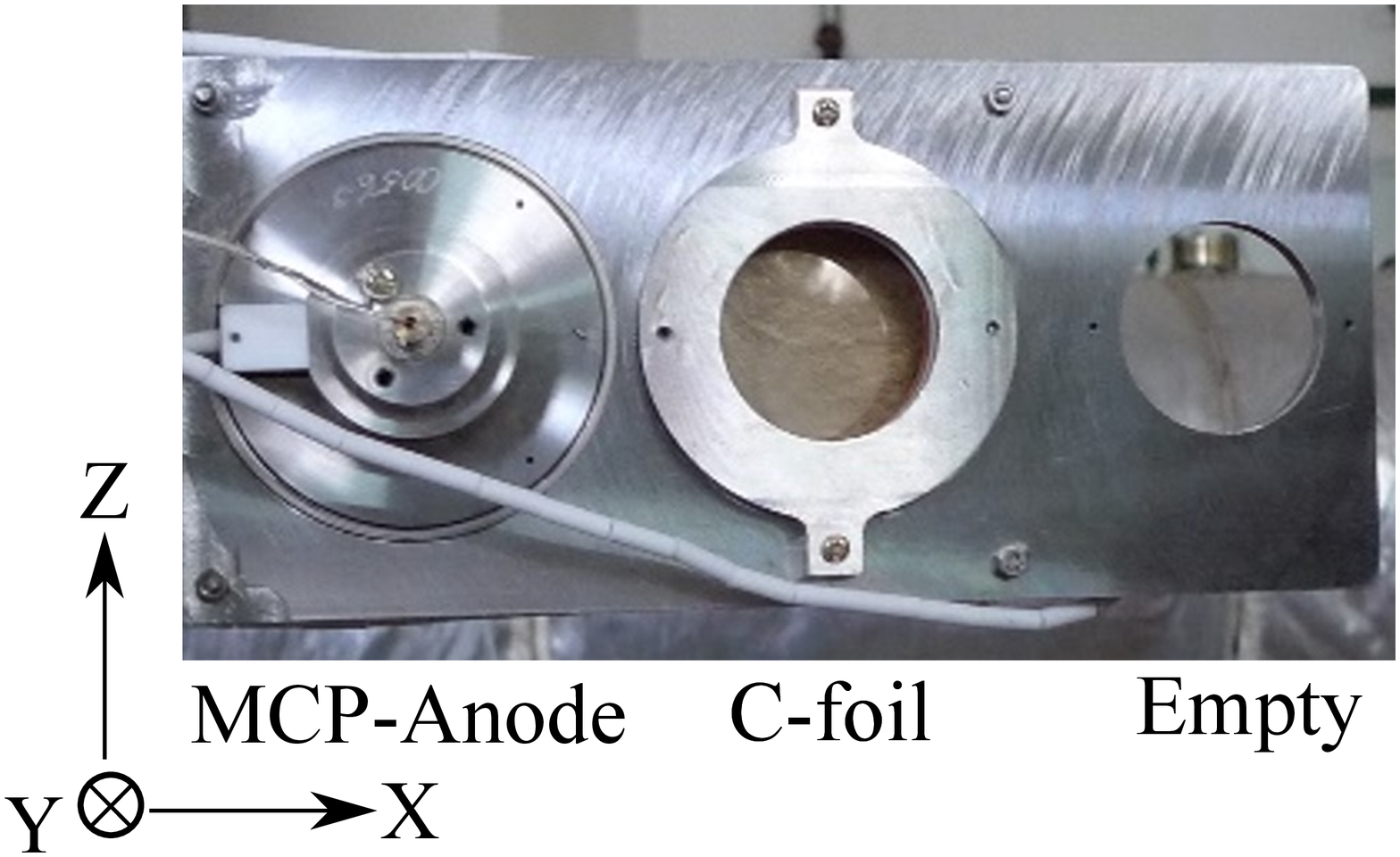}
			\includegraphics[width=.55\hsize]{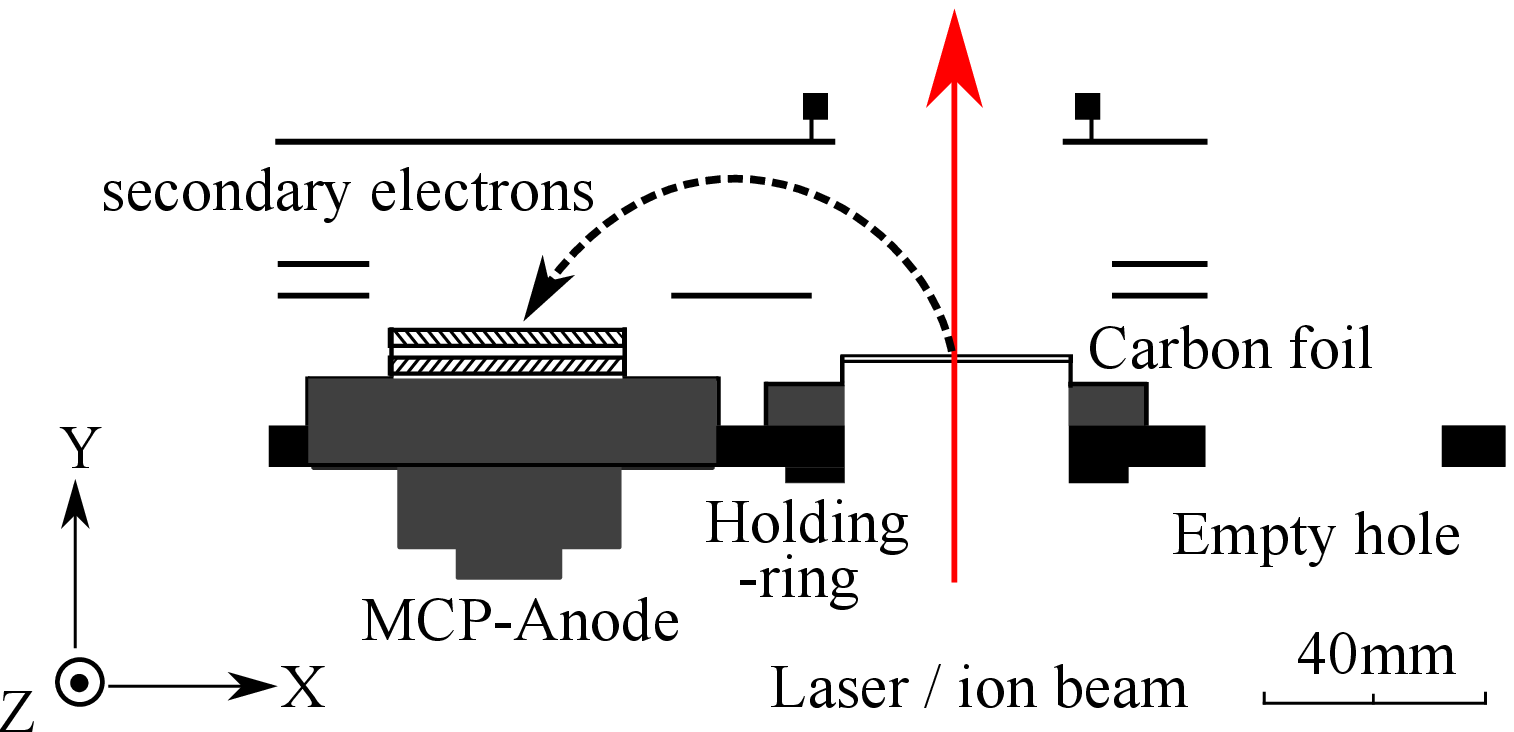}
		\end{center}
		\caption{(Color online) Front and schematic view of the Time-Of-Flight detector \cite{tu2010,mei,zhang14}. The ion/laser beam pass through a thin carbon foil in Y-direction and release secondary electrons which are isochronously transported to the MCP detector placed in forward direction. The detector is movable between experimental and testing chambers in X-direction (see Fig. \ref{fig1}). The angle of 
			the 658 nm laser beam can be scanned to probe the 3D positions of the objects using diffusely reflected laser light from surface of the carbon foil or other materials.}
		\label{TOFdetector}
	\end{figure}

	\section{Principle}\label{sec.2}
	
	A schematic view of the CSRe storage ring is shown in Fig. \ref{fig1}. With the two TOF detectors installed in the dipole-free straight section of the storage ring, the velocity of the stored ions (laser light) can be measured as: 
	\begin{equation}
	v = \frac{L}{t_{2}-t_{1}+\Delta t_{delay1-2}},
	\label{vFormula}
	\end{equation}
	where $L$ is the distance between the two detectors which is approximately 18 meters, $t_1$ and $t_2$ are the arrival times of the ions (laser light) at the two detectors which are registered by the oscilloscope, $\Delta t_{delay1-2}$ is the time delay difference which comes with signal cable length difference between the two TOF detectors (see Fig. \ref{fig1}). The structure of the TOF detector is shown in Fig. \ref{TOFdetector}. When ions/laser beam penetrates the carbon foil of the TOF detector, secondary electrons (SEs) are released from the carbon foil. The electrons are guided to the Micro-Channel Plates (MCPs) by applied electric and magnetic fields. The electrical signals formed from the MCP anode are transmitted through signal cable to the oscilloscope and are recorded as timing signals for off-line analysis. External trigger either from the kicker of CSRe or a pulsed Laser instrument is used to activate the data recording of the digital oscilloscope. Surely there is a time delay between the arrival time of the TOF signal which is registered by the oscilloscope and the actual arrival time of ions (laser light) at the carbon foil of the TOF detector. Although the electric and magnetic field settings, the cable length are identical for two TOF detectors, the time delays are not exactly the same. Here we use $\Delta t_{delay1-2}=t_{delay1}-t_{delay2}$ to indicate the time delay difference.  
	
	For accurate velocity measurement of the stored ions in the ring, the two parameters ($L$ and $\Delta t_{delay1-2}$) should be calibrated. It is possible to use the primary beam with well-defined velocity, which could be extracted from the synchrotron CSRm, sent through the beam-transport line and directly injected to CSRe \cite{jxia}, for the velocity calibration of the cocktail secondary beam as long as the ions of both primary and secondary beam undergo the same $L$ in the straight section of CSRe. However, the velocity spread of the primary beam is about $\sigma(v)/v=2\times10^{-4}$ \cite{jxia,mao16}. There is also a risk to use the primary beam because the TOF detectors can only handle at maximum hundreds of ions stored simultaneously in CSRe. Therefore, an independent way to determine $L$ and $\Delta t_{delay1-2}$ by using two types of laser beams has been developed. The method reported here is not only useful for the velocity measurements in the present and future storage ring facilities but can also be applied in other similar general-purpose experiments.

	\subsection{Precision goal and the error budgets}\label{sec.2.1}
	
	In a typical IMS experiment \cite{ZhangPRL2012}, tens of heavy ions selected from projectile fragments of the primary beam are injected and stored simultaneously in the ring. The velocities of the stored ions are between 0.66$c$ to 0.73$c$ depending on the mass-to-charge ratios of the ions. Therefore $t_2-t_1$ ranges from about 82 ns to 92 ns for stored ions and is about 60 ns for a laser beam. 
	The precision of velocity measurement
	is determined by the precision of time measurements, variations of $L$ and of $\Delta t_{delay1-2}$:
	\begin{equation}
	\frac{\sigma(v)}{v} =\sqrt{ \frac{\sigma^2(L)}{L^2}+\frac{\sigma^2(t)}{t^2}},
	\label{dvFormula}
	\end{equation}
	where $t=t_{2}-t_{1}+\Delta t_{delay1-2}$. 
	The goal \cite{xu15,shuai16} of the velocity measurement is to reach a precision of $\sigma(\bar{v})/\bar{v}\approx2\times10^{-5}$, where $\bar{v}$ is the mean velocity owing to the fact that the velocity of the stored ions is gradually reduced due to energy loss in the TOF detectors. With this precision, we hope to reach a maximum mass resolving power of $R\approx8\times10^{5}(1\sigma)$ assuming $\sigma(\gamma_t)/\gamma_t\approx1\times10^{-4}$ could be reached in the data analysis \cite{chen18} while the acceptance of $\sigma(B\rho)/(B\rho)\approx8.5\times10^{-4}$ was preserved in CSRe.
	
	The time precision $\sigma(t)$ is limited by the time-resolution of the TOF detector which is about 18.5 ps \cite{zhang14} and can be statistically improved by measuring $t$ for more than one time (recording timing signals for hundreds of revolutions) for each ion.
	For $\sigma(L)$, the situation is a little different.
	Although no dipole magnets are installed in the straight section, there are quadrupole magnets in between the two TOF detectors. So, the resulting trajectories of the ions at the straight section of the ring will obey betatron oscillations around `closed' orbits. Therefore, the ions pass through the TOF detector at different positions on the foil and these positions vary from one revolution to another. It is found by simulations \cite{ruijiu15}, that by assuming the carbon-foils in the TOF detectors were perpendicularly aligned to the ion beam, the betatron oscillation will cause a variation in trajectory of $dL/L\approx2\times10^{-6}$. However, if the carbon foils were not flat but supposedly have a surface roughness of 0.18 mm, this would introduce an  additional error of $dL/L=1\times10^{-5}$. Moreover, if the surfaces of the two carbon foils are not parallel to each other, $L$ and $t$ might be correlated and the measured $v$ might be biased if we assume a constant $L$ in the data analysis. Therefore a scan of the exact values of $L$ between the two ultra-thin carbon foils are needed. Furthermore, although the $\Delta t_{delay2-1}$ could be considered as constant during a IMS experiment, its value should be calibrated anyhow especially if the connection signal cables of the TOF detectors were changed before the experiment. 
	
	In the following chapters, the $L$ and $\Delta t_{delay1-2}$ measurements employing a 658 nm laser range finder and a short-pulsed ultra-violet (UV) laser will be described and the distance and timing properties of the double TOF detector system in CSRe will be characterized in detail.
	
	\section{Distance measurement}\label{sect_distance_measurement}
	\subsection{Method}\label{sec.II}
	\begin{table}
		\begin{center}
			\caption{Main parameters of the laser range finder MS60 MultiStation \cite{ms60}.}
			\label{parMS60}       
			\begin{tabular}{l@{ }c@{ }c@{$\ $}c@{$\ $}c@{$\ $}c@{$\ $ }c@{$\ $ }c@{$\ $ }c@{$\ $ }c@{$\ $ }c@{$\ $ }c@{$\ $}c@{$\ $}c@{$\ $ }c@{$\ $ }c@{$\ $ }c@{$\ $ }c}
				\noalign{\smallskip}\hline
				Measurement technology:   &    Wave Form Digitizing \\ 
				Wave length:   &    658 nm             \\
				Pulse duration: &    1.5 ns (FWHM)                  \\
				Repetition rate: &  2 MHz             (scan mode) \\
				&  4 MHz             (point mode) \\
				Spot size at 20 m:   &  3.3 mm$\times$8 mm             \\
				Angle precision  : &  1$^{\circ}$/3600 (ISO 177123-3) \\
				Absolute distance precision:   &  2 mm + 2ppm (ISO 177123-4) \\
				\noalign{\smallskip}\hline
			\end{tabular}
		\end{center}
	\end{table}
	The objects of the distance measurement are two carbon-foils of the TOF detectors with their centers aligned to the ion beam trajectory. 
	Schematic view and a picture of the TOF detector \cite{tu2010,mei,zhang14} are shown in Fig. \ref{TOFdetector}. The diameter and thickness of the foils are 40 mm and $\approx$ 90 nm (20 $\mu$g/cm$^2$), respectively. Both TOF detectors are moveable inside the vacuum chamber in X direction in such a way that the carbon foil can be placed in and out of the beam line. Usually, each TOF detector is prepared and assigned in a separate vacuum chamber, a testing chamber which is connected to the experimental chamber with a valve gate (see Fig. \ref{fig1}).
	The difficulties of distance measurements here are the non-reachable environment which is inside an ultra-high vacuum chamber ($\approx10^{-11}$ mbar)
	and the fragility of the studied objects (90 nm thick carbon foils), therefore traditional laser tracker technology using reflection minors \cite{cai16} is not applicable.

	The task is accomplished by using a 658 nm laser range-finder Leica Nova MS60 MultiStation (hereafter MultiStation) \cite{ms60}.  
	Typical parameters of the MultiStation are listed in Table \ref{parMS60}.
	\begin{figure}
		\begin{center}
			\includegraphics[width=.59\hsize]{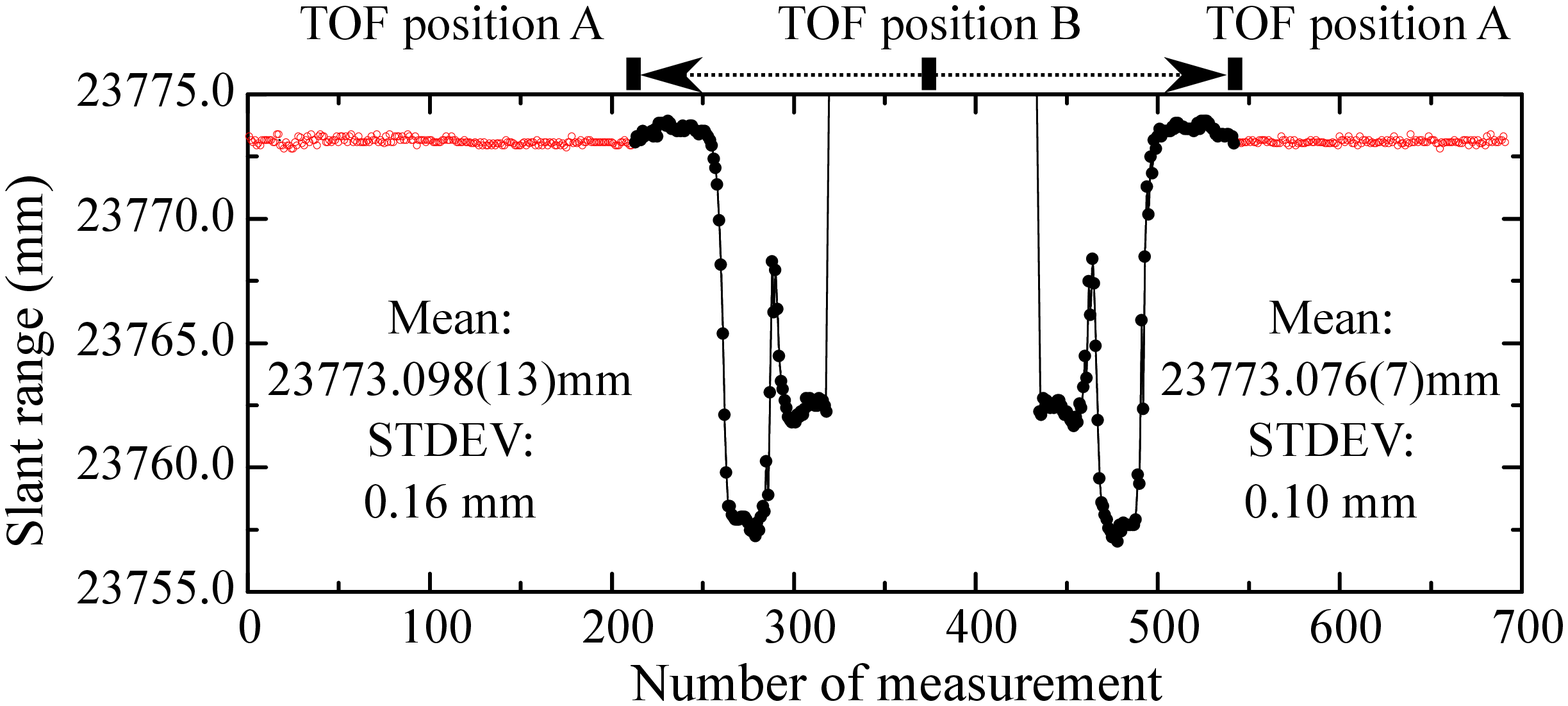}
		\end{center}\label{pos_scan}
		\caption{(Color online) Measured distances between TOF2 detector and MS60 MultiStation. The laser beam direction of the MultiStation was fixed and the TOF2 detector was moved in X direction from position A to B and back. Position A (B) indicates the center of the carbon foil (empty hole) was on the laser beam. Red open circles indicate the distance measured when TOF2 detector was stationary placed in position A. Black closed circles indicate the distanced measured when TOF2 detector was moving: the change of measured distances was observed when the holding ring and mounting frame of the foil or empty hole on a frame was on the path of laser beam (see Fig. \ref{TOFdetector}). }
		\label{fig3}
	\end{figure}
	During the measurements, a pulsed laser beam from the MultiStation was sent through the vacuum 
	beam pipe and directly to the carbon foil of the TOF detector (see Fig. \ref{fig1}). The light from the laser beam was diffusely reflected from the carbon foil and detected by the MultiStation. The time-of-flight of the reflected light pulse is used to determine the distance 
	from the carbon foil to the MultiStation. The errors \cite{iso} of the distance measurement by the MS60 MultiStation consist of two parts: a zero-point correction (offset) of 2 mm and a statistical relative error of 2 ppm (which means an error of 0.04 mm for a range of 20 m) \cite{ms60}.
	As we are interested in the distance between the two TOF detectors, that is the range difference between the two detectors, the influences of the laser passing forward and backward through air and a quartz window of the vacuum chamber as well as the zero-point corrections of the MultiStation cancel out.
	
	Two working modes, the point mode and the scan mode, were used in the distance measurements. 
	In the point mode, the measurements were repeated at a fixed angle of the laser beam. In the scan mode, the angle of the laser beam was scanned through a selected area of the object and only one distance value was recorded for each angle. The distance between TOF1 and TOF2 was measured as slant range difference ($l_2-l_1$) in the point mode and as distance difference in Y direction ($Y_2-Y_1$) in the scan mode (see Fig. \ref{TOFdetector}). 
	
	\subsection{Preparation of the measurements}\label{sec3.2}
	\begin{figure*}[htb!]
		\begin{center}
			\includegraphics[width=0.8\hsize]{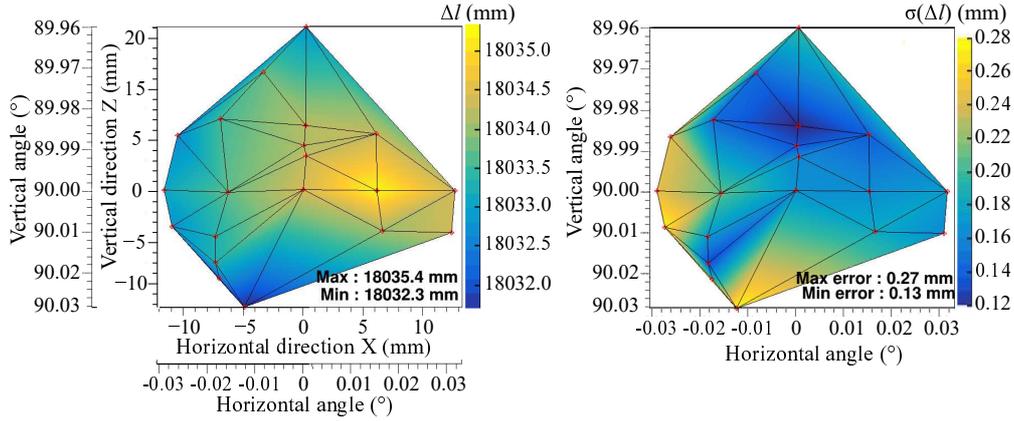}
			\caption{(Color online) Results obtained in the point mode. Left: the distance ($\Delta l$) between two TOF detectors as a function of horizontal and vertical angles. The laser beam pointing at (Hz,V)=$(0^{\circ},90^{\circ})$ angle was aligned to the beam line. The corresponding X and Z positions of the laser spot on the carbon foil of TOF2 detector are also indicated in additional axes. The position (X,Z)=(0,0) corresponding to the centers of the two carbon foils (see text in Chapter 3.2). Right: The corresponding width ($\sigma(\Delta l)$) of the slant distance distribution. The experimental results are represented by red points and are color-coded (see legend). The variation of $\Delta l$ reached 3 mm at maximum. Note that there are areas on the foils where no distance can be measured. }\label{fig4}
		\end{center}
	\end{figure*}
	The laser beam of the MultiStation has to be aligned with the center of the beam line before the measurements. A collimation axis aligned with the beam line center was assigned by the accelerator department and its extension was marked as a collimation line on the floor.
	The MultiStation was set on a tripod and the tripod was positioned in a place where the light spot ($\Phi=2.5 mm$) of the build-in Laser plummet of the MultiStation was on this collimation line on the floor. After the position of the tripod was fixed, the MultiStation was leveled with the aid of a build-in electronic level device. This would ensure that the laser beam is on the horizontal plane when the vertical angle read is 90 degree. With the aid of a build-in collimation camera of the MultiStation, the horizontal and vertical angles of the two carbon-foil centers were measured by the MultiStation after the TOF detectors were moved in to the center of the beam line. The location and height of the tripod shall be adjusted iteratively until the laser beam is aligned with the beam line center. For each adjustment of the tripod, the MultiStation must be leveled again. The criteria for  alignment were: the horizontal angles of the two foil-centers were the same within $\pm0.005$ degree and the corresponding two vertical angles were 90 degree within $\pm0.005$ degree. After the criteria were fulfilled, the laser beam pointing to the center of carbon foil of the TOF2 detector was assumed to be aligned with the beam line when the vertical angle read in the MultiStation was 90 degree. The corresponding horizontal angle was then set to be zero and a Cartesian coordinate system was set in the MultiStation such that X=Z=0 corresponding to vertical angle of 90 degree and horizontal angle of zero. All measurement results were expressed in this coordinate system.
	
	During the preparation, the accuracy and reliability of the measurements were tested by monitoring the distance of TOF2 detector using the point measurement mode. At first the laser beam was shot on the center of the carbon foil and the data acquisition (DAQ) rate of the MultiStation was set to be 1 Hz. The results of the distance measurements are shown with red open circles on the left-hand side of Fig. \ref{fig3} (position A). The standard deviation ($\sigma$) of the measured data is 0.16 mm. Afterwards the detector was moved out from the beam position (position A) along X direction to the empty hole position on the foil frame (position B) (see Fig. 2) and reverse. The data were taken while the detector was moving (black filled circles). When the holding ring, mounting frame and carbon-foil of TOF2 detector were on the beam path the distances were measured as 23757.9 mm, 23762.1 and 23773.1 mm respectively. The relative distances between different parts of the detector agree with the machining and the assembly of the detector, supporting the reliability of the measurements. Finally, the detector was moved back to position A and similar distance was measured as before with a mean value of 23773.1 mm and a standard deviation of 0.1 mm. This indicates the reliability of the moving mechanisms of the detectors.
	
	\subsection{Measurement results}\label{sec.III}
	\begin{figure*}[htb!]
		\includegraphics[width=\hsize]{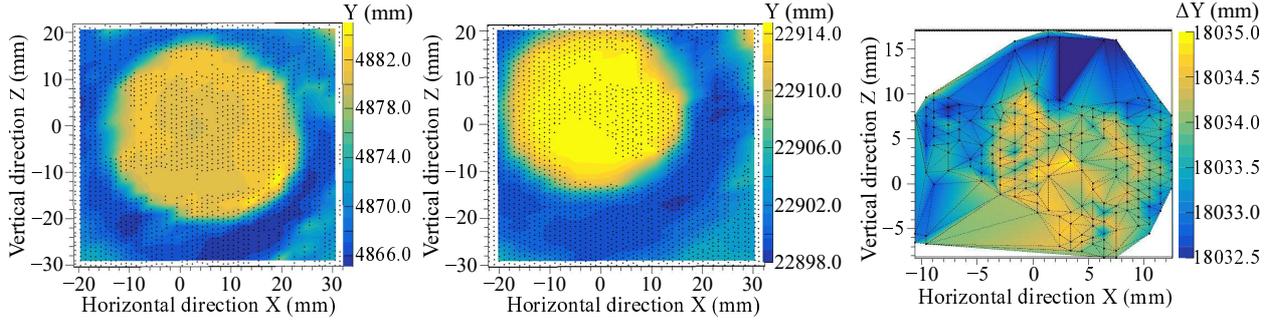}
		\caption{(Color online) Results obtained in the scan mode. Left and middle panels: the 3D positions of two ultra-thin carbon foils of TOF1 and TOF2 detectors inside the vacuum pipe. The Cartesian coordinate system is defined by the MS60 MultiStation at the beginning of the scanning, where Y axis is aligned to the beam direction of the ions. The experimental results are represented by black points and the color-code of the Y-axis is given in the legend. The ring and mounting frame which are closer to the MultiStation than the carbon-foil are clearly seen. There are areas in ultra-thin foils where no data points could be measured. Right panel: The point-to-point distance ($\Delta$Y) between the two carbon-foils.\added{ The source data for figures as well as photographs of the two carbon foils can be downloaded at \cite{Yan2018}}}\label{fig5}
	\end{figure*}
	
	In Fig. \ref{fig4}, the results of the point mode measurements for 20 different angles are summarized. First at a certain angle of the laser beam, TOF1 detector was removed from the beam line and the slant range ($l_2$) of TOF2 detector was measured for about 400 times. Then, TOF1 detector was moved into the beam-line and the slant range ($l_1$) was measured for about 400 times. The distance between the two TOF detectors was calculated as the difference of the mean slant ranges: $\Delta l=\bar{l_2}-\bar{l_1}$. The width of the distance distribution was given by $\sigma(\Delta l)=\sqrt{\sigma^2(l_1)+\sigma^2(l_2)}$. The measurements were repeated for different angles to deduce the distance between two TOF detectors over the whole areas of the two carbon foils. It turned out that only a limited area on the carbon foils were measurable: the diffusely reflected light of the laser pulses from certain areas was too weak to be detected by the MultiStation and no slant range data could be returned for certain angles of the laser beam. Beside that, the measured $\Delta l$ ranged from 18032.3 mm to 18035.4 mm (see left-hand side of Fig. \ref{fig4}) and $\sigma(\Delta l)$ at each angle of the laser beam varied from 0.13 mm to 0.27 mm (see right-hand side of Fig. \ref{fig4}). The maximum distance variation of 3 mm was observed. The disadvantage of the point mode measurements was the fact that the TOF1 detector should be moved back and forth for each angle of the laser beam. 
	
	3D-positions of the two carbon foils obtained in the scan mode measurements are shown in Fig. \ref{fig5}. The angle of the laser was varied automatically in pre-set steps to scan once for all the whole area of the carbon foil. After TOF1 detector was scanned, it was moved out from the beam line to let TOF2 detector to be scanned. The scanning raster for each detector was 1 mm in both X and Z directions and the DAQ rate is 62 Hz. Only one distance data was returned for each angle in the scan mode as compared to hundreds of distance data for each angle in the point mode.
	In Fig. \ref{fig5}, there are also some areas in the ultra-thin carbon foils where no data points could be measured in the scan mode, which is consistent with the results obtained in the point mode (see Fig. \ref{fig4}). Nevertheless, the positions of the carbon foils and their holding rings are clearly seen in the color-coded maps in Fig. \ref{fig5}. A tilting angle of 2.3$^{\circ}$ between the two carbon foils was measured and the surface roughness of the TOF1 and TOF2 carbon foils were measured to be at maximum 1.3 mm and 0.6 mm, respectively. In comparison, the measured surface roughness of the holding rings in the detectors were only about 0.3 mm. The centers of the two foils were turned out to be not in the same altitude: the carbon foil of TOF2 was located about 6.7 mm higher than the carbon foil of TOF1, which means that the two foils overlap by only 79\% of the areas from the beam point of view. Moreover, on the XZ-plane, the overlapped area that were measured was less than 20$\times$20 mm$^2$, mainly restricted by the smaller measurable area of the ultra-thin carbon foil in TOF2 detector. Along Y direction, the point-to-point distances ($\Delta Y=Y_2-Y_1$) of the two carbon foils had a mean value of 18033.85 mm and were ranging from 18032.5 to 18035.0 mm, agreed with the results of the point mode measurements. The distance variations had a standard deviation of $\sigma(\Delta Y)=$0.6 mm and the main cause of the variations was the roughness of the foil surfaces, especially the surface of the carbon foil in TOF1 detector.
	
	\section{Time delay difference measurement}\label{section_time delay difference measurement}
	\subsection{Method}
	Given the measured distance between the two TOF detectors, the time delay difference $\Delta t_{delay1-2}$ between the two TOF detectors was measured with help of a dedicated  ultra-violet (UV) laser with 15 ps (FWHM) pulse duration. The main parameters of the UV laser are summarized in Table \ref{UltraViolerLaser}.
	The selection of the UV laser was based on following considerations:
	(1) the energy of photon shall be equal to or higher than the work function of the ultra-thin carbon foils (about 4 - 5 eV \cite{cwf,she2013}) to induce electron emissions from the foil surface, here 266 nm wavelength corresponding to a photon energy of 4.66 eV;
	(2) the pulse duration should be as short as possible, at least shorter than 18 ps which is the time resolution of the TOF detector \cite{zhang14};
	(3) the power of laser beam should be lower than 20 mW (measured at 1 MHz) to avoid burning the carbon foils;
	(4) the divergence angle of laser beam should be as small as possible.
	\begin{table}\begin{center}
			\caption{The main parameters of the ultra-violet laser FL-266nm-Pico \cite{fl266}.}	\label{UltraViolerLaser}       
			\begin{tabular}{l@{ }c@{ }c@{$\ $}c@{$\ $}c@{$\ $}c@{$\ $ }c@{$\ $ }c@{$\ $ }c@{$\ $ }c@{$\ $ }c@{$\ $ }c@{$\ $}c@{$\ $}c@{$\ $ }c@{$\ $ }c@{$\ $ }c@{$\ $ }c}		\noalign{\smallskip}\hline\noalign{\smallskip}
				Wave length:   &    266.456 nm             \\
				Pulse duration: &    15 ps (FWHM)                  \\
				Pulse energy:   &    10 nJ                  \\
				Repetition rate: &  1-1000 Hz              \\
				Spot size (90\%) at 20 m:   &  8.2 mm$\times$7.2 mm             \\		\noalign{\smallskip}\hline
		\end{tabular}\end{center}
	\end{table}
	
	The experimental set-up is shown in Fig. \ref{fig1}. The UV laser instrument was placed on an optical anti-shake platform nearby TOF2 detector. Two fused-silica broadband reflecting mirrors was used to reflect and guide the UV laser towards the center of the beam line. Since the UV laser light was very dim and no telescope and camera functions were built in the UV laser instrument, the 658 nm visible laser light of the MS60 MultiStation was used to guide the optical path for the UV laser. The 658 nm laser beam was first aligned with the center of the beam line as described in chapter \ref{sec3.2}, sent through the quartz window at the other end of the vacuum beam pipe and shot onto the two reflecting minors (see Fig. \ref{fig1}). The orientations of the two minors were adjusted to reflect the 658 nm laser onto the light outlet of the UV laser instrument. The position and direction of the UV laser instrument was adjusted to allow the UV laser light be transmitted backwards along the path of the 658 nm laser light to the centers of two carbon foils in TOF2 and TOF1 detectors. The laser from the MultiStation was turned off after the UV laser was aligned with the beam line center and ready for use in the time delay difference measurements.
	
	Every 0.1 second, a UV laser pulse was shot through TOF2 to TOF1 detector (see Fig. \ref{fig1}). A trigger signal from the UV laser instrument was sent to the oscilloscope (type: \replaced{LeCory WaveRunner 604Zi}{DPO 71254c}) to activate data recording for each shot. The laser pulse produced electric signals at the TOF detectors by releasing secondary electrons from the carbon foil surfaces (see Fig. \ref{TOFdetector}). The timing signals from MCP anodes were transmitted along signal cables from two TOF detectors to the two input channels of the oscilloscope and were recorded for off-line analysis. The sampling rate and the voltage resolution of the oscilloscope were \replaced{10}{50} GHz and \replaced{0.57}{0.8} mV, respectively. The measured $\Delta t_{delay2-1}$ is given by:
	\begin{equation}
	\Delta t_{delay1-2} =( t_{1}-t_{2})-\frac{L}{c},
	\label{DeltaTFormula}
	\end{equation}
	where $c = 299 792 458 $ m/s and $L = 18033.85 $ mm $\pm$ 1.25 mm, $t_{1}$, $t_{2}$ are arrival times of the TOF1 and TOF2 timing signals registered by the oscilloscope. 
	
	\begin{figure}[htb!]
		\begin{center}	\includegraphics[width=0.58\hsize]{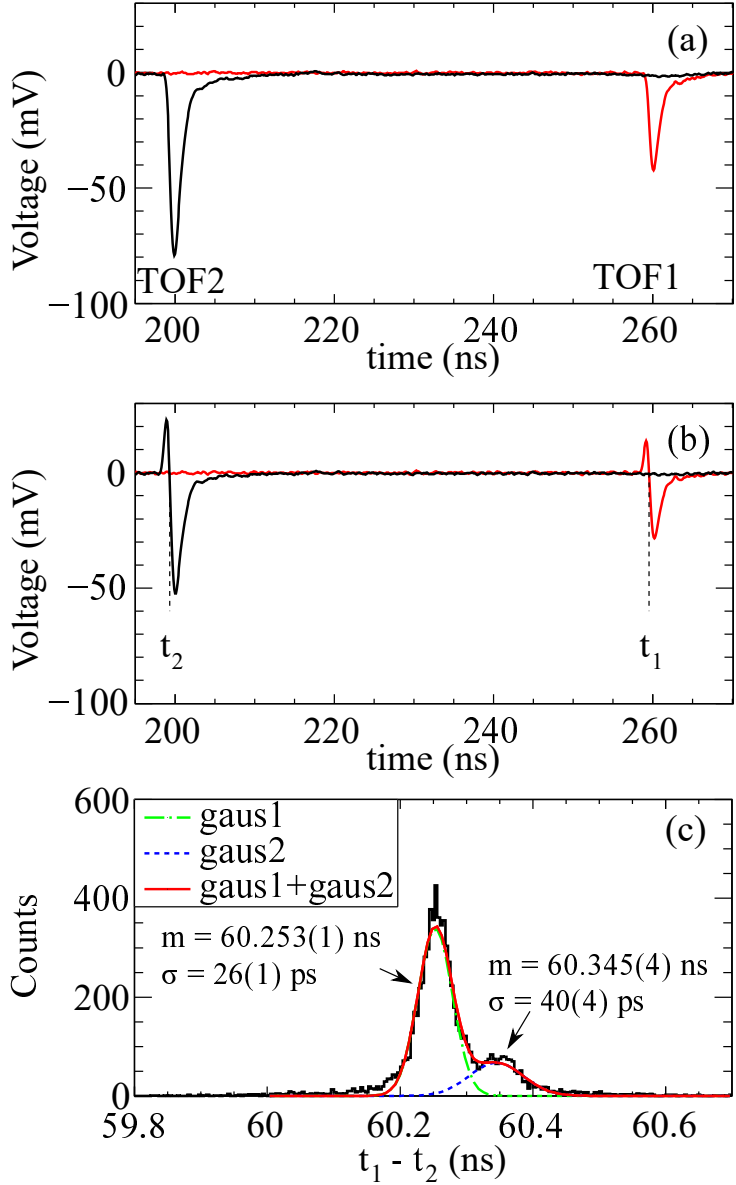}
		\end{center}
		\caption{(Color online) (a) Typical timing signal spectra registered by the oscilloscope for one laser pulse.
			The black and red lines are timing signals from TOF2 and TOF1 detector, respectively. (b) The corresponding signal processing by using a standard Constant Fraction technique.
			$t_{1}$ and $t_{2}$ are the registered arrival times of the laser pulse at TOF1 and TOF2 detector.
			(c) A distribution of time difference between the two TOF detectors measured for 8000 laser pulses. A combination of two Gaussian functions is used to fit the histogram. The fit result is shown as solid red line. The two corresponding decomposed Gaussian functions are shown as green and blue dashed lines.
		}
		\label{result}
	\end{figure}
	
	\subsection{Results and discussion}
	\begin{figure}[htb!]
		\begin{center}
			\includegraphics[width=0.85\hsize]{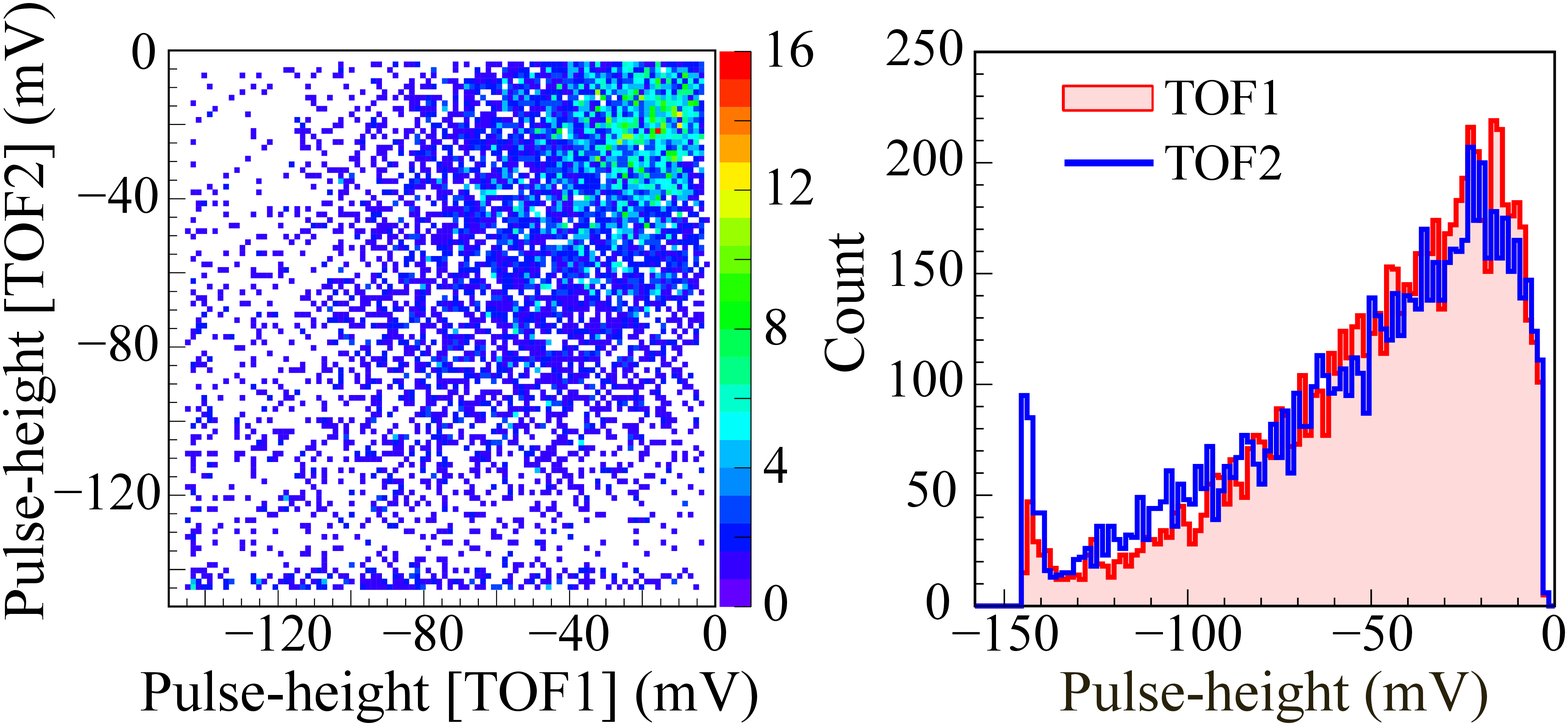}
		\end{center}
		\caption{(Color online) Left: 2D histogram of the pulse-heights of the pair of the timing signals induced by UV laser pluses. The detection efficiency of the UV laser pluses is 100\%. Right: 1D projections of the 2D histogram. The pulse-height of timing signal followed a Landau-like distribution.
		}
		\label{fig7}
	\end{figure}

	In total 8000 pairs of timing signals induced by 8000 shots of the UV laser pluses were recorded. The detection efficiency of the 266 nm pulsed laser light was 100\% for both TOF1 and TOF2 detectors.
	
	Typical timing signal spectra for one laser pulse is shown in Fig. \ref{result}(a). The standard Constant Fraction timing discrimination (CF) process of the two timing signals is shown in Fig. \ref{result}(b). The arrival times of signals ($t_1$, $t_2$) were determined at
	the zero-crossing points as shown in Fig. \ref{result}(b).
	The time difference $t_1-t_2$ was calculated for each laser pulse.
	The distribution of all the 8000 $t_1-t_2$ values is shown in Fig. \ref{result}(c) where a double peak feature was observed. A combined function composed of two Gaussian functions was used to fit the distribution histogram.
	The two decomposed Gaussian functions deduced from the fitting are shown as green and blue dashed lines in Fig. \ref{result}(c).
	The main peak located at 60.253 ns has a distribution $\sigma$ = 26 ps and the side peak located at 60.345 ns has $\sigma$ = 40 ps. Indications from the manufacturer of the FL-266nm-Pico laser suggested that the side peak could originated from 
	the time structure of the laser pulse. Improvement of the pulse time structure of the laser shall be done in the future. 
	
	The time-of-flight between the two TOF detector for the laser light pulse in vacuum is $L/c=60.154 $ ns $\pm $ 4 ps as inferred from distance measurements. Therefore, the time delay difference is $\Delta t_{delay1-2}= 99(26)$ ps using the results of the main peak in Fig. \ref{result}(c). Note that, the time resolution
	of one TOF detector can be estimated to be about $\sqrt{(26^2-4^2)/2}=18$ ps assuming the two detectors had the same time resolution. This result agrees well with our previous 
	off-line test result of $\sigma = 18.5 (2)$  ps using alpha particles \cite{zhang14}.
	
\begin{figure}[htb!]
		\begin{center}
			\includegraphics[width= 0.85\hsize]{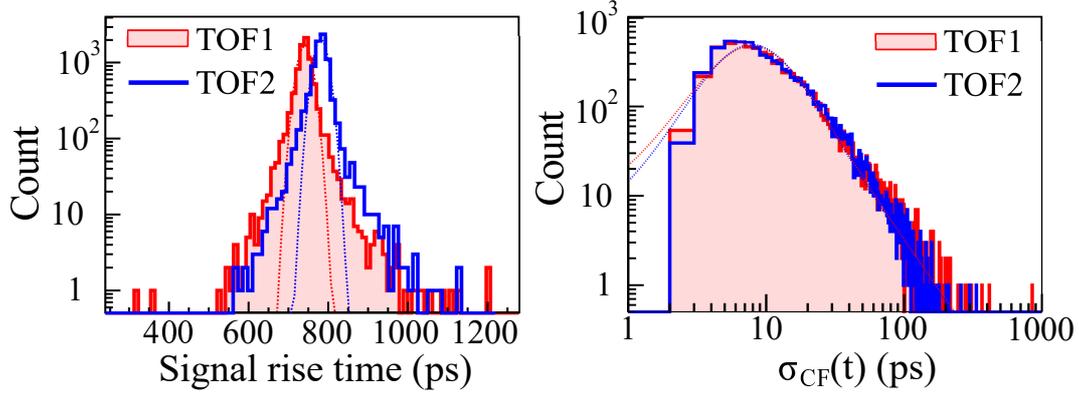}	
		\end{center}
		\caption{(Color online) Left: The signal rise time (from 10\% to 90\% pulse-height) distributions and the corresponding Gaussian fitting results (dotted lines). The mean values of the distributions are 741.1(2) ps for signals from TOF1 detector and 781.5(2) ps for TOF2. The sigma values of the distributions are 17.2(2) ps for TOF1 signals and 16.7(2) ps for TOF2 signals. Right: The distributions of the Constant Fraction (CF) timing errors of the signals and the corresponding Landau fitting results (dotted lines). The fitted most probable value (MVP) of the $\sigma_{CF}(t)$ distribution is \replaced{8.5(1) ps for both TOF1 and}{14.2(1) ps for TOF1 and 14.4(1) ps for} TOF2.
		}
		\label{result2}
	\end{figure}

	The recorded timing signals from the two detectors tended to have same pulse-height and the pulse-height followed a Landau-like distribution as shown in Fig. \ref{fig7}. The signal rise time (from 10\% to 90\% pulse-height) distributions for the timing signals are shown on the left-hand side of Fig. \ref{result2}. The signal rise times have a mean value of 741/782 ps and a distribution $\sigma$ of 17/17 ps for TOF1/TOF2 detector respectively. The CF timing errors $\sigma_{CF}(t)$, originated from uncertainties of the voltage recording (\replaced{0.57}{0.8} mV) and finite sampling rate (\replaced{10}{50} GHz) of the oscilloscope, are shown on the right-hand side of Fig. \ref{result2}. The most probable value (MVP) of the $\sigma_{CF}(t)$ distribution is \replaced{8}{14} ps for both TOF detectors, which is consistent with the estimated time resolution of 18 ps of one TOF detector.

	\section{Summary and outlook} \label{sec.IV}
	
	The distance between the two ultra-thin carbon foils and the time delay difference of two TOF detectors under ultra-high vacuum condition
	were measured \emph{in-situ} employing two types of lasers. With a 658 nm laser range finder MS60 MultiStation, the distance has been measured to be 18033.85 mm with a variation of $\pm 1.25$ mm 
	within a measurable overlapped $20\times20$ mm$^2$ area between the two carbon foils. The result from two measuring modes, the point and scan modes, are consistent with each other. Revealed by the scanning results, the main contribution of the distance variations comes from the surface 
	roughness of the two foils which is on the level of 1 mm. The tilting angle between the two foils was found to be about 2.3 degree which means a transverse distance of 20 mm on the carbon foils will introduce 20 mm$\times$ tan$(2.3^{\circ})$ = 0.8 mm difference to the flight path between the two TOF detectors in straight section.
	The determined distance variation of $dL=1.25$ mm due to surface roughness of the carbon foils translates into a relative uncertainty of 
	$dL/L\approx6.9\times10^{-5}$. 
	
	The time delay difference has been measured to be $\Delta t_{delay1-2}$ = 99(26) ps with the help of a pulsed 266 nm laser with 15 ps (FWHM) pulse duration. The time resolution of the double TOF detector system translated into $\sigma(t)/t= 26 \text{ ps} /60514 \text{ ps}=4.3\times10^{-4}$ for the time-of-flight of light between TOF1 and TOF2 detectors. The deduced time resolution of one TOF detector is about 18 ps, agrees well with our previous measurement. The short pulsed UV laser is approved to be a convenient tool for testing the performance of TOF detectors. Note that the tilting angle between the two carbon foils may bring in an additional error of $\sigma(t)\approx4$ ps to the $t_1-t_2$ of the stored ions. Considering the UV laser only illuminated a small part (5\%) of the detector ($\Phi=$ 40 mm diameter of carbon foil compared to 9 mm$\times$8 mm laser spot at TOF1) while each stored ion can pass through the detector at any point of its active area, the time resolution of the double TOF detector system for the stored ions will be worse than $\sigma(t)/t=3.1\times10^{-4}$ for ions with a velocity of $0.73c$.
	
	In conclusion, the current velocity-resolution of the double TOF detector system at CSRe is about $\sigma(v)/v\approx4.4\times 10^{-4}$ for the speed of light, mainly limited by the time-resolution of the TOF detectors. For the stored ions in CSRe, $\sigma(v)/v$ ranges from $2.9\times 10^{-4}$ to $3.2\times 10^{-4}$ depending on the velocity of the stored ions. In order to reach the aimed statistical precision of $\sigma(\bar{v})/\bar{v}=2\times 10^{-5}$ for the measured mean velocity \cite{xu15,shuai16}, the distance variations should be reduced to the level of $d(L)/L=2\times10^{-5}$ (i.e.\ 0.36 mm/18 m) and at least two hundred revolutions are required for the circulating ions in CSRe. However, after passing 400 times through the 20 $\mu$g/cm$^2$ carbon foils, the velocity of a chromium ion (Z $=$ 24) will be reduced by about $\Delta v/v=2.4\times10^{-4}$. Thinner carbon foils are suggested for the design of new TOF detectors. It is clear that the research and development works of new TOF detectors with better timing resolution are crucial in achieving better velocity-resolution and thus higher mass resolving power of IMS at the current and future storage ring facilities.
	
	Many aspects, concerning the future improvements, were revealed in this work. Firstly, the altitudes of the two carbon-foils were different to each other by 6.7 mm which would limit the acceptance of the ring. Secondly, 
	the tilting angle between the two foils was measured to be about 2.3 degree which would introduce additional error in the velocity measurement. Thirdly, the surface roughness of the two ultra-thin carbon foils was on the level of 1 mm and the main contribution to the distance variations. New carbon foils with better surface flatness are needed. Replacement of old carbon foils and assembly adjustments of the altitude and angle of the two TOF detectors are planned.
	In general, laser range-finders are useful position monitoring tools for any fragile objects in place with complicated access, such as TOF detectors inside a vacuum pipe. Some areas of the ultra-thin carbon foils could not be measured by the Leica MS60 MultiStation in this work. The higher power of the carrier laser could be considered to overcome this problem. Mechanical and assembly improvements of the TOF detectors could be better monitored using such a custom-tailored MultiStation. Furthermore, the pulse time structure of the FL-266nm-Pico laser should be tested and improved in future measurements. The spot sizes of the UV laser beam at TOF2 and TOF1 were about 2.0 mm$\times$1.7 mm and 9.4 mm$\times$8.2 mm, respectively. This difference might cause the distribution broadening of $t_1-t_2$ in Fig. \ref{result}(c). A pulsed laser beam with smaller divergence angle is recommended for the future measurements.
	
	The measurement scheme developed in this paper is planned to be integrated into the data acquisition system of the IMS experiment for on-line velocity calibrations of the stored ions. In IMS experiments at CSRe, pulsed ion beams are injected into the ring every 20 seconds and the TOF measurements are done within 1 millisecond after injection. Pulsed UV laser beam will be shot onto the two TOF detectors in between every injection of ion beam for the determination of $\Delta t_{delay1-2}$ value. Before and immediately after the IMS experiment, the $L$ parameter will be measured using the MS60 MultiStation.
	
	For future storage ring at HIAF facility \cite{yang}, two TOF detectors equipped with ultra-thin carbon foils of even larger diameters are foreseen to be installed at a larger distance apart in a straight section of the Spectrometer Ring (SRing) \cite{bwu}. Therefore it is suggested to install quartz vacuum windows in the beam line to allow diagnosis with laser beams. The same idea can be used for the ILIMA project at the future FAIR facility in Germany \cite{ILIMA2006,bosch11,pwalker}, where two TOF detectors are to be installed in the Collector Ring CR \cite{Dolinskii2008}.

	\section*{Acknowledgment}
	Enlightening discussions with Prof. Klaus Blaum, Dr. Liyan Zhang, Dr. Yun Liu, Dr. Yuan Liu, Dr. Jie Yang and technical supports from Leica Geosystems company are greatly acknowledged.
	This work is supported in part by 
	the NSFC (Grant Nos. 11605248, 11605252, 11711540016), National Key R$\&$D Program of China (Contract No. 2016YFA0400504 and No. 2018YFA0404400), the Key Research Program of the Chinese Academy of Sciences (No. XDPB09),
	the Helmholtz-CAS Joint Research Group HCJRG-108,
	the External Cooperation Program of the CAS (GJHZ1305), the European Research Council (ERC) under the European Union's Horizon 2020 research and innovation programme (grant agreement No 682841 "ASTRUm") and the U.S. Department of Energy, Office of Science, Office of Nuclear Physics, under contract No. DE-AC-06CH11357. X.Y. acknowledge partial support by the FRIB-CSC Fellowship under Grant No. 201704910964. R.C. acknowledge support by the Office of China Postdoctoral Council. We are grateful to the referees whose remarks lead to substantial improvements in the writing of this paper.
	
\end{document}